\documentclass[aps,pra,twocolumn,showpacs,amssymb]{revtex4}

\usepackage{graphicx}
\usepackage{dcolumn}
\usepackage{bm}

\begin{document}
\title{Trirefringence in nonlinear metamaterials}
\author{Vitorio A. \surname{De Lorenci}}
 \email{delorenci@unifei.edu.br}
\author{Jonas P. Pereira$\mbox{}^{*\ddagger}$}
 \email{jonaspedro.pereira@gmail.com}
 \affiliation{$\mbox{}^*$Instituto de Ci\^encias Exatas,
Universidade Federal de Itajub\'a, Itajub\'a, MG 37500-903, Brazil}
\affiliation{$\mbox{}^{\ddagger}$ Universit\'e de Nice Sophia Antipolis, Nice, CEDEX 2, Grand Chateau Parc Valrose}
\affiliation{$\mbox{}^{\dag}$Dipartamento di Fisica and ICRA, Sapienza Universit\`{a} di Roma, P.le Aldo Moro 5, I-00185 Rome, Italy}

\date{\today}

\begin{abstract}
We study the propagation of electromagnetic waves in the limit of geometrical optics
for a class of nearly transparent nonlinear uniaxial metamaterials for which their
permittivity tensors present a negative principal
component. Their permeability are assumed positive and dependent on the electric field.
We show that light waves experience triple refraction
-- trirefringence. Additionally to the ordinary wave, two extraordinary waves
propagate in such media.
\end{abstract}

\maketitle

\section{Introduction}
%
Electromagnetic waves in a material medium propagate according to Maxwell's
equations complemented by certain relations linking strengths and induced fields --
the constitutive relations.
Depending on the dielectric properties of the medium
and also on the presence of applied external fields, a variety of
optical effects can be found. One of wide interest is the birefringence
\cite{Landau,born,paschotta2008},
occurring when electromagnetic waves propagate in media exhibiting two distinct
refractive indexes \cite{Landau} in a same wave vector direction.
This effect occurs naturally in some well--known crystalline solids, as quartz and 
sapphire for instance. In nonlinear media, where the dielectric coefficients are 
field dependent, birefringence could also be induced by the presence of external 
fields, leading to the known Kerr, Cotton--Mouton and magnetoelectric effects
\cite{roth, delorenci2008}.
Birefringence has been widely used in technology of optical devices \cite{paschotta2008} and
as a powerful experimental technique for investigating properties of physical systems,
including biological \cite{bio} and astrophysical ones \cite{astro1,astro2}, among others.
The phenomenon of triple refraction has been much less investigated.
By trirefringence in a given wave vector direction, we mean the existence of
three distinct refractive indexes in that direction.
In the realm of linear electrodynamics, trirefringence does not occur \cite{wood1969}.
Nevertheless, nothing was studied when
the dielectric tensors are field dependent (nonlinear electrodynamics).

Multirefringent properties have been measured in tailored photonic crystals
\cite{tri1}. Such materials are constructed to manipulate light propagation and hence
can lead to a variety of applications \cite{Joannopoulos}.
However, in these media, structural details (lattice constants, defects, etc) are
imperative and hence they can not be described in terms of effective dielectric tensors \cite{smith2}.
Developments in photonic band gap materials and the so-called
metamaterials have enabled the discovery of several new
phenomena \cite{smith2}. For instance, it was experimentally shown
\cite{smith} that nearly transparent isotropic metamaterials allow light propagation
when both effective dielectric coefficients (permittivity and permeability)
are negative.
In fact, this phenomenon and other unusual properties displayed by isotropic
media  with negative dielectric coefficients were long ago proposed
theoretically \cite{veselago}.
It is worth emphasizing
these media must be dispersive and the negative coefficients are obtained for
convenient frequency ranges.
Some other unusual properties exhibited by specific metamaterials are
negative refractive index \cite{shelby},
trapping of light \cite{trapping,smolyaninov3}, perfect lens
devices \cite{pendry2}, the electromagnetic cloaking effect
\cite{cloak1,smolyaninov2} and the occurrence of asymmetry for
propagation of light in  opposite wave vector directions \cite{asymmetric}.
Wave propagation in indefinite
metamaterials (where not all the principal components of the
dielectric tensors have the same sign) has been also considered
\cite{smith2}, showing that effects already proposed in the context
of isotropic metamaterials can be obtained and possibly improved.
Indefinite  metamaterials can also be used for investigating
certain aspects of General Relativity \cite{smolyaninov}.

In this paper, we show that nonlinear metamaterials described in
terms of effective dielectric tensors, may display trirefringence. Analytical
expressions describing this effect are formally obtained from
Maxwell's electromagnetism and a simple theoretical model
is numerically examined. A possible experimental realization of the
media expected to display this effect is also addressed.
The vectorial three dimensional formalism \cite{delorenci2010} is used.
The units are set such that $c=1$.

In the next section electromagnetic wave propagation in
nonlinear materials is examined. The eigenvalue problem is stated
and formally solved for a class of nonlinear materials presenting
non-isotropic but constant permittivity tensor (for a given frequency) and
isotropic permeability dependent on the modulus of the resultant electric field.
In Sec. III, trirefringence phenomenon is theoretically described
and proposed to occur in nearly transparent nonlinear uniaxial
metamaterials. The analysis is performed in terms of phase and group velocities.
Final remarks and conclusions are presented in Sec. IV.

\section{Wave propagation}
The electrodynamics of a continuum medium at rest in the absence of
sources is governed by the Maxwell field equations
\begin{eqnarray}
\vec{\nabla}\cdot \vec{D}\,&=&\,0,\;\;\;\;\;
\vec{\nabla}\times \vec{E}\,=\,-\frac{\partial \vec{B}}{\partial t},
\label{1a}
\\
\vec{\nabla}\cdot \vec{B}\,&=&\,0,\;\;\;\;\;
\vec{\nabla}\times \vec{H}\,=\,\frac{\partial \vec{D}}{\partial t},
\label{1b}
\end{eqnarray}
taken together with the constitutive relations between the fundamental
fields $\vec{E}$ and $\vec{B}$, and the induced ones $\vec{D}$ and $\vec{H}$,
written here as
\begin{equation}
D_i = \sum_{j=1}^3\varepsilon_{ij}E_j,\;\;\;
H_i = \sum_{j=1}^3\mu_{ij}B_j.
\end{equation}
The dielectric coefficients $\varepsilon_{ij}$ and  $\mu_{ij}$
are the components of the permittivity and the inverse permeability tensors,
respectively, and they encompass all information about the electromagnetic properties of the
medium. Further, for any vector $\vec{\alpha}$ we denote its $i$th component by $\alpha_i$ ($i=1,2,3$).

The propagation of monochromatic electromagnetic waves is here examined within
the limit of geometrical optics \cite{Landau} using the method of field
disturbances \cite{Hadamard}. This method can be summarized as follows \cite{delorenci2010}.
Let $\Sigma$, defined by $\phi(t,\,\vec{x})=0$, be a smooth (differentiable of
class ${\cal C}^n, n>2$) hypersurface. The function $\phi$ is understood
to be a real-valued smooth function of the coordinates ($t,\vec{x}$) and regular
in a neighborhood $U$ of $\Sigma$. The spacetime is divided by $\Sigma$ into
two disjoint regions $U^-$, for which $\phi(t,\,\vec{x})<0$, and $U^+$,
corresponding to $\phi(t,\,\vec{x})>0$.
The discontinuity of an arbitrary function  $f(t,\,\vec{x})$ (supposed to be
a smooth function in the interior of $U^\pm$) on $\Sigma$ is a smooth
function in $U$, and is given by \cite{Hadamard}
\begin{equation}
\label{discontinuity}
\left[f(t,\,\vec{x})\right]_{\Sigma}
\doteq \lim_{\{P^\pm\}\rightarrow P}
\left[f(P^+) - f(P^-)\right],
\end{equation}
with $P^+,\,P^-$ and $P$ belonging to $U^+,\,U^-$ and $\Sigma$,
respectively.
The electromagnetic fields are supposed to be smooth functions in the interior
of $U^+$ and $U^-$ and continuous across $\Sigma$ ($\phi$ is now taken as the eikonal
\cite{Landau} of the wave). However they
have a discontinuity in their first derivatives, such that \cite{Hadamard}
\begin{eqnarray}
{\left[\partial_{t}E_i\right]}_{\Sigma}&\!\!\!=\!\!&\omega e_i\,,\;\;\;\;\;\;
{\left[\partial_{t}B_i\right]}_{\Sigma}=\omega b_i\,,
\label{q1}\\
{\left[\partial_i E_j \right]}_{\Sigma}&\!=\!&-q_i e_j,\;\;\;
{\left[\partial_i B_j \right]}_{\Sigma}\,\!=\!-q_i b_j,
\label{q2}
\end{eqnarray}
where $e_i$ and $b_i$ are related to the derivatives of the electric and magnetic
fields on $\Sigma$, and correspond to the components of the polarization of the propagating
waves \cite{delorenci2002}. The quantities $\omega$ and $q_i$ are the angular frequency and
the {\it i}th component of the wave vector.
(Incidentally, we note that the negative signs appearing in Eq. (\ref{q2}) are missing in the
corresponding equations in Ref. \cite{delorenci2010}.)

For the cases of interest in this work, the permittivity of the media under study will be described by real
diagonal tensors (losses have been neglected) whose components
$\varepsilon_{ij}$
are dependent only upon the constant frequency of the wave. We set the magnetic permeability of these media
to be real functions of the modulus of the
electric field, such that 
\begin{equation}
\mu_{ij}(|\vec{E}|)=\frac{\delta_{ij}}{\mu(|\vec{E}|)},
\end{equation} 
where $\delta_{ij} = {\rm diag}\,(1,1,1)$.
Thus, applying the boundary conditions stated by Eqs. (\ref{q1}-\ref{q2}) to the field
equations (\ref{1a}-\ref{1b}) we obtain the eigenvalue equation \cite{delorenci2004,delorenci2008}
\begin{equation}
\displaystyle\sum_{j=1}^{3} Z_{ij}\; e_{j}=0,
\label{1}
\end{equation}
where the Fresnel matrix $Z_{ij}$ is given by
\begin{eqnarray}
Z_{ij} = \varepsilon_{ij} - \frac{\mu'}{\omega\mu^2}(\vec{q}\times\vec{B})_{i}E_{j}
-\frac{1}{\mu\omega^2}\left(q^2 \delta_{ij} - q_{i} q_{j}\right),
\label{16b}
\end{eqnarray}
with
\begin{equation}
\mu' \doteq \frac{1}{|\vec{E}|}\frac{\partial \mu}{\partial |\vec{E}|}
\end{equation}
and $q^2 \doteq \vec{q}\cdot\vec{q}.$

Non-trivial solutions for the eigenvalue problem in Eq. (\ref{1}) can
be found if, and only if, $\det \mid Z_{ij}\mid \,=\, 0$, which is known as
the generalized Fresnel equation. This equation gives also the dispersion
relations of the media under study. Using the covariant formulae for the traces
of linear operators \cite{rodrigues} the Fresnel equation can be cast as
\begin{eqnarray}
\det\,\mid Z_{ij}\mid\,=
-\frac{1}{6}({Z_{1}})^{3}\,
+\,\frac{1}{2}{Z_{1}}{Z_{2}}\,-\frac{1}{3}\,{Z_{3}}=0,
\label{23}
\end{eqnarray}
where we defined the traces
\begin{eqnarray}
Z_{1} &\doteq& \sum_{i=1}^{3} Z_{ii},
\label{tracesa}
\\
Z_{2} &\doteq& \sum_{i,j=1}^{3} Z_{ij}\,Z_{ji},
\label{tracesb}
\\
Z_{3} &\doteq& \sum_{i,j,l=1}^{3}Z_{ij}\,Z_{jl}\,Z_{li}.
\label{tracesc}
\end{eqnarray}
As a requirement of the geometrical optics limit, the wave fields
are considered to be negligible when compared with the external
fields. Thus, we assume from now on that the fields are approximated by their
external counterparts $\vec{E}_{ext}$ and $ \vec{B}_{ext}$.
We set
$\vec{E} \approx \vec{E}_{ext} \doteq E\,\hat x$
and
$\vec{B} \approx \vec{B}_{ext} \doteq B\,\hat y,$
which could be arbitrary functions of space and time
coordinates. Let us examine the par\-ticu\-lar case of uniaxial media
\cite{Landau, born} with permittivity
\begin{equation}
\varepsilon_{ij}={\rm diag}\,(\varepsilon_{\scriptscriptstyle\parallel},
\,\varepsilon_{\scriptscriptstyle\perp},\,\varepsilon_{\scriptscriptstyle\perp}).
\end{equation}
Using Eqs. (\ref{16b}) and (\ref{tracesa}-\ref{tracesc}), straightforward calculations show that Eq. (\ref{23})
results in the following algebraic fourth degree equation for the phase velocity $v = \omega/q $
of the propagating waves,
\begin{equation}
a\,v^4+b\,v^3+c\,v^2+d\,v+e=0
\label{40},
\end{equation}
with
\begin{eqnarray}
 &a =& 6\,\varepsilon_{\scriptscriptstyle\perp}{}^2\,\varepsilon_{\scriptscriptstyle\parallel},
\label{47}\\
 &b =& \frac{6\,\mu'}{\mu^2}\,\varepsilon_{\scriptscriptstyle\perp}{}^2\, E\, B\,
\hat{q}_z,
\label{55}\\
 &c =& -\frac{6}{\mu}\,\varepsilon_{\scriptscriptstyle\perp}\left[
2\varepsilon_{\scriptscriptstyle\parallel}\,\hat{q}_x{}^2
+ (\varepsilon_{\scriptscriptstyle\perp}
+ \varepsilon_{\scriptscriptstyle\parallel})(\hat{q}_y{}^2+\hat{q}_{z}{}^2)\right],
\label{56}\\
 &d =& -\frac{6\,\mu'}{\mu^3}\,
\varepsilon_{\scriptscriptstyle\perp}\,E\,B\,\hat{q}_z,
\label{60}\\
 &e =& \frac{6}{\mu^2}
\left[\varepsilon_{\scriptscriptstyle\parallel}\,\hat{q}_x{}^2
+\varepsilon_{\scriptscriptstyle\perp}\,(\hat{q}_y{}^2+\hat{q}_z{}^2)\right].
\label{61}
\end{eqnarray}
We defined $\hat{q}_\alpha \doteq (\hat q \cdot \hat \alpha)$,
where $\hat{q} \doteq \vec{q}/q$, for any unit
vector $\hat \alpha$. Then, $\hat{q} = \hat{q}_x\hat{x} +
\hat{q}_y\hat{y} +  \hat{q}_z\hat{z}$ and $\hat{q}^2 \doteq \hat{q}\cdot \hat{q} = 1$.
Notice that when the propagation occurs in
the $xy$-plane (spanned by the external electric and magnetic fields),
the coefficients $b$ and $d$ are null and the generalized Fresnel equation
reduces to a quadratic equation in $v^2$, therefore allowing only birefringence.
The same behavior occurs if  $\mu' = 0$. In fact in this situation we recover
linear electrodynamics.

Solving Eq. (\ref{40}) we obtain
%
\begin{equation}
v_o = \pm\, \frac{1}{\sqrt{\mu\,\varepsilon_{\scriptscriptstyle\perp}}}
\label{75},
\end{equation}
\begin{equation}
v^{\scriptscriptstyle\pm}_e
=-\sigma \hat{q}_z \pm\, \sqrt{(\sigma\hat{q}_z)^2 + \frac{1}{\mu
\varepsilon_{\scriptscriptstyle\parallel}}
\left(\frac{\varepsilon_{\scriptscriptstyle\parallel}}
{\varepsilon_{\scriptscriptstyle\perp}}\hat{q}_x^2+\hat{q}_y^2+\hat{q}_z^2\right)},
\label{76}
\end{equation}
where
\begin{equation}
\sigma \doteq \frac{\mu'\,E\,B}{2\mu^2\varepsilon_{\scriptscriptstyle\parallel}}.
\label{76ib}
\end{equation}
The solution $v_o$ does not depend on direction of the wave propagation and
will be called the ordinary wave, whereas $v^{\scriptscriptstyle\pm}_e$
depend on direction of wave propagation and will be called
extraordinary waves \cite{Landau}.
By definition, the velocities
of the waves are given by $\vec{v} = v\hat{q}$, where $v$ is given by Eqs. (\ref{75}-\ref{76}).
In order to achieve more
simplicity in the following analysis we assume the external fields to be constant and set the wave vector in a given direction of the
$xz$-plane, i.e., $\hat{q}_y = 0$, $\hat{q}_x = \sin\theta$ and
$\hat{q}_z = \cos\theta$. In this notation $\theta$ indicates the angle
between $\hat{q}$ and  the $z$ direction.

\section{Trirefringence}
Two distinct solutions for $\vec{v}^{\scriptscriptstyle\pm}_e$ in a same given direction $\hat{q}$
can be obtained from Eq. (\ref{76}) if
\begin{equation}
-1 <\frac{1}{\mu\,\varepsilon_{\scriptscriptstyle\parallel}\,\sigma^2}
\left(\frac{\varepsilon_{\scriptscriptstyle\parallel}}
{\varepsilon_{\scriptscriptstyle\perp}}\tan^2\theta + 1 \right) < 0.
\label{84}
\end{equation}
Thus, taking into account the ordinary wave, Eq. (\ref{84}) defines a region
inside which trirefringence occurs in any chosen direction $\hat{q}$.
Let us examine this effect closer.
\begin{figure}[!hbt]
\leavevmode
\centering
\includegraphics[scale = 1.05]{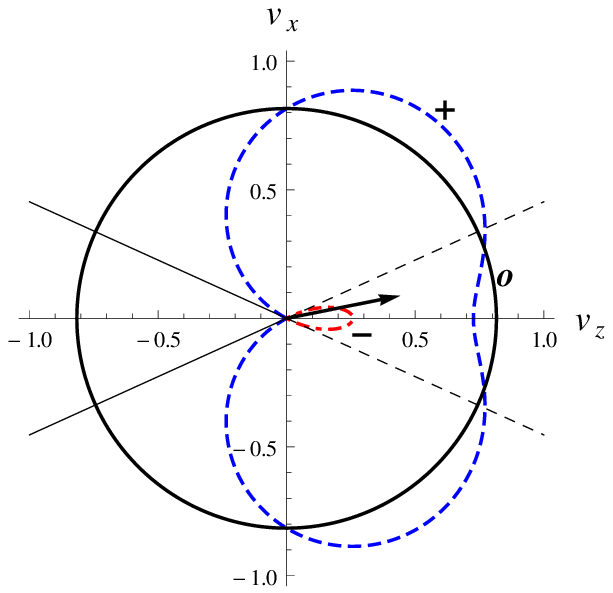}
\caption{{\small\sf (color online). Normal surfaces \cite{Landau, born} of
a nonlinear medium with dielectric coefficients given by $\varepsilon_{ij}=
{\rm diag}\,(-\epsilon_{\scriptscriptstyle\parallel},
\,\varepsilon_{\scriptscriptstyle\perp},\,\varepsilon_{\scriptscriptstyle\perp})$ and  $\mu(E)$.
We set a model where the numerical values were taken such that
$\epsilon_{\scriptscriptstyle\parallel}=6.69$, $\varepsilon_{\scriptscriptstyle\perp}
= 1.88$, $\mu = 0.8$ and $\mu'E\,B = 4.21$.
The ordinary wave is represented by the circular solid line and the extraordinary waves
are represented by the dashed and dot-dashed curves.
The dashed straight lines encompass an angular region in which trirefringence occurs.
In this case there exist two extraordinary waves denoted by $+$ and $-$. The symbol $o$
refers to the ordinary wave.}}
\label{fig1}
\end{figure}
In order to guarantee the existence of an ordinary wave we must set
$\mu\varepsilon_{\scriptscriptstyle\perp} > 0$, otherwise $v_o$ is not
real. This is true when both coefficients $\mu$ and $\varepsilon_{\scriptscriptstyle\perp}$
present the same sign, which can be positive for
usual media or negative for left handed materials (in this case a negative
refractive index occurs \cite{veselago,shelby,smith3}). Let us set these coefficients
to be positive. Now, in order to satisfy Eq. (\ref{84}) we set
$\varepsilon_{\scriptscriptstyle\parallel} =
-\epsilon_{\scriptscriptstyle\parallel} < 0$. Hence, trirefringence
occurs in directions determined by
\begin{equation}
\frac{\varepsilon_{\scriptscriptstyle\perp}}{\epsilon_{\scriptscriptstyle\parallel}}
> \tan^2\theta
> \frac{\varepsilon_{\scriptscriptstyle\perp}}{\epsilon_{\scriptscriptstyle\parallel}}
\left(1-\epsilon_{\scriptscriptstyle\parallel}\,\mu\sigma^2\right).
\label{angles}
\end{equation}
The above discussed phenomenon is displayed in Fig. (\ref{fig1}) where the
normal surfaces \cite{Landau,born} associated with the ordinary and
extraordinary waves are
depicted for some specific values of the quantities appearing in Eq. (\ref{angles}).
For any given direction encompassed by the angles between the two dashed
straight lines, defined by Eq. (\ref{angles}), there are three distinct
solutions: the dashed and dot-dashed curves representing the extraordinary waves,
and the circular solid curve representing the ordinary wave.
For angles between the dashed and the solid straight lines
only birefringence occurs, and finally only one refraction occurs for directions
encompassed by the angles between the two solid straight lines.
It is also worth noticing from Fig. (\ref{fig1}) that in the sectors where more then one
refractive index occur, the medium under consideration behaves as a positive
or negative medium \cite{Landau, born}, depending on sub-sectors and extraordinary waves.

In geometrical optics the directions of light rays are
given by the directions of the group velocities
\begin{equation}
\vec{u} = \frac{\partial\omega}{\partial\vec{q}}
= v\hat{q} + q \frac{\partial v}{\partial\vec{q}}
\label{groupvs}
\end{equation}
which are considered as the physical velocities of propagation of the rays \cite{Landau}.
As we see from Eq. (\ref{76}), the extraordinary waves depend on the wave vector.
Thus, the directions of the extraordinary light rays do not in general coincide with
the directions of the extraordinary phase velocities, as explicitly shown by Eq. (\ref{groupvs}).
Taking $v=\{v_o,v_e^+,v_e^-\}$ into the definition of the group velocity
we obtain that the ordinary group velocity is identified with the ordinary phase
velocity $\vec{u}_o = v_o\hat{q}$.
Nevertheless, the extraordinary group velocities are given by
\begin{equation}
\vec{u_e} = u_x \hat{x} + u_z \hat{z},
\label{group}
\end{equation}
with
\begin{eqnarray}
u_{x} = \frac{v_e^3\sin\theta + \sigma v_e^2 \sin 2\theta + \frac{v_e}
{\mu\varepsilon_{\scriptscriptstyle\parallel}}
\!\!\left(\!\frac{\varepsilon_{\scriptscriptstyle\parallel}}
{\varepsilon_{\scriptscriptstyle\perp}}-1\!\right)\sin\theta\cos^2\!\theta}
{v_e^2 + (1 - \eta)\sigma v_e\cos\theta +
\frac{\eta}{2\mu\varepsilon_{\scriptscriptstyle\parallel}}\cos^2\theta},
\label{groupx}
\\
u_{z} = \frac{v_e^3\cos\theta + \sigma v_e^2 \cos 2\theta -
\frac{v_e}{\mu\varepsilon_{\scriptscriptstyle\parallel}}
\!\!\left(\!\frac{\varepsilon_{\scriptscriptstyle\parallel}}
{\varepsilon_{\scriptscriptstyle\perp}}-1\!\right)\cos\theta\sin^2\!\theta}
{v_e^2 + (1 - \eta)\sigma v_e\cos\theta +
\frac{\eta}{2\mu\varepsilon_{\scriptscriptstyle\parallel}}\cos^2\theta},
\label{groupz}
\end{eqnarray}
and where we defined
\begin{equation}
\eta \doteq \frac{\omega}{\varepsilon_{\scriptscriptstyle\parallel}}
\frac{\partial\varepsilon_{\scriptscriptstyle\parallel}}{\partial\omega}.
\end{equation}
\begin{figure}[!bt]
\leavevmode
\centering
\includegraphics[scale = 1.05]{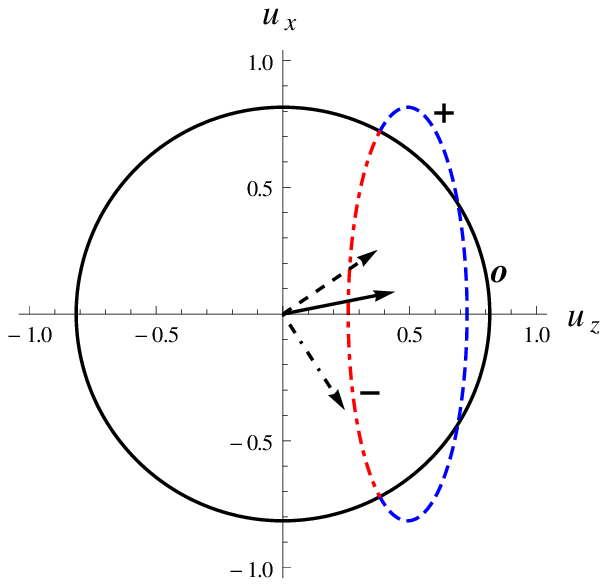}
\caption{{\small\sf (color online). Ray surfaces \cite{Landau,born} of the nonlinear medium
considered in Fig. \ref{fig1}.
The ordinary group velocity is represented by the circular solid line and
the extraordinary group velocities are represented by the dashed and dot-dashed curves.
}}
\label{fig2}
\end{figure}
In these equations $v_e\doteq v_e^{\pm}$ represents the extraordinary wave velocities in the direction $\hat{q}$,
which can be obtained from Fig. \ref{fig1} by taking the distances from the associated
points on its wave vector surfaces to the origin of the coordinate system.
For the particular model set in Fig. \ref{fig1} the corresponding ray surfaces
\cite{Landau,born} are depicted in Fig. \ref{fig2} as a function of
the angle $\varphi$ between $\vec{u}$ and the z axis, where it was assumed for simplicity that $|\eta| << 1$. Notice that this assumption resides
in the realm of nonlinear media. In the
context of linear electrodynamics, for which $\mu'=0$, Eqs. (\ref{group}-\ref{groupz})
show that $|\eta| << 1$ leads to superluminal group velocity solutions.
Therefore, this assumption is not in disagreement with the so-called causality requirement
\cite{smith2}.
For the ordinary waves we have $\varphi=\theta$, as anticipated.
For the extraordinary waves, the relationship between these angles can be
obtained directly from Eqs. (\ref{groupx}-\ref{groupz})
as $\varphi = \arctan [u_x(\theta)/u_z(\theta)]$.
From Fig. \ref{fig2}, we see that there is a sector inside which there exist two
extraordinary rays in any chosen direction of observation.
For the complementary sector there are none extraordinary rays.
We point out that any chosen direction for the rays inside the trirefringent sector
is related to two different extraordinary wave vectors in Fig \ref{fig1}.
Similarly, for any given wave vector inside the two dashed straight lines in
Fig. \ref{fig1}, there are two different extraordinary rays associated
with it in Fig \ref{fig2}.
For instance, let us take the wave vector in the direction $\theta = \pi /16$,
as indicated by the arrow drawn in Fig. \ref{fig1}.
As is clear, in this direction there are three
solutions for the phase velocity, indicating that trirefringence occurs.
These solutions are represented by the points on the $+$, $-$ and $o$ curves,
found in this particular direction, and
their moduli are $v_e^+=0.76$, $v_e^-=0.20$ and $v_o=0.82$,
respectively. From Eqs. (\ref{group}-\ref{groupz}), the
corresponding group velocities present magnitudes
$u_e^+=0.83$, $u_e^-=0.55$ and $u_o=0.82$. Their associated directions
are indicated by the arrows appearing in Fig. \ref{fig2}, where the dashed and
dot-dashed arrows correspond to $\vec{u}_e^+$ and $\vec{u}_e^-$, respectively.

\section{Final remarks}
%
If the principal permittivity components are also dependent upon the
modulus of the electric field, then analogous calculations leading to the equations for the phase and group
velocities and conditions for having trirefringence derived in this paper will still hold by identifying
\begin{equation}
\varepsilon_{\scriptscriptstyle\parallel} \rightarrow \varepsilon_{\scriptscriptstyle\parallel}(E)
+ E\,\frac{\partial \varepsilon_{\scriptscriptstyle\parallel}(E)}{\partial E},
\end{equation}
and
\begin{equation}
\varepsilon_{\scriptscriptstyle\perp} \rightarrow \varepsilon_{\scriptscriptstyle\perp}(E).
\end{equation}

Once we work in the regime of high frequencies (required by geometrical optics) losses
must be assumed to be low \cite{Landau}.
Metamaterials with low-losses are still a matter of current
investigations \cite{low_loss1}, with some achievements already obtained in such regime \cite{low_loss2}.

In what concerns the role played by the nonlinearities Eqs. (\ref{76}-\ref{76ib})
show us that if they exist (i.e. $\mu'\neq0$), irrespective of their strengths, then
they will lead to the presence of two extraordinary waves (hence allowing trirefringence).
Given the physical quantities appearing in Eqs. (\ref{group}-\ref{groupz}), the thresholds for
the strengths of the nonlinearities are such that causality is not violated.
The important quantity related to the nonlinearities is $\sigma$, given by Eq. (\ref{76ib}),
which is experimentally controllable, thereby making the predictions here derived feasible.

Layered media \cite{pendry_layered,born} seem good candidates for experimental
realizations of our assumptions. Consider a system constituted by
a repetition of two thin low-loss layers. One of them is composed by a nonmagnetic
($\mu_1=1$) and dispersive ($\epsilon_1=\epsilon_1(\omega)$) medium. The other one
is composed of a liquid medium whose permeability and permittivity are
dependent upon the modulus of the resultant electric field as
$\mu_2= 1- f(|\vec{E}|)$ and $\epsilon_2=1- g(|\vec{E}|)$, where $f(|\vec{E}|)$ and
$g(|\vec{E}|)$ are usually much smaller than unity \cite{baranova}.
Additionally, if convenient external fields and layer parameters are present, then it is always possible
to guarantee that our constraints are fulfilled.

Refraction analyzes were disregarded in our work once they are just a straightforward
extension of the analyzes valid for birefringent crystals \cite{born}.
The only difference now is the existence of two extraordinary waves.

Summing up, working in the limit of geometrical optics we studied the propagation of electromagnetic
waves in nearly transparent nonlinear uniaxial metamaterials in the presence of external electric
and magnetic fields.
We assumed a constant non-isotropic permittivity tensor (for a given frequency) presenting a negative principal component
and an isotropic permeability dependent on the modulus of the electric field. We solved the corresponding eigenvalue
problem and obtained the general forth degree polynomial equation, whose solutions describe the propagation of
waves. In this context, we showed that trirefringence is a phenomenon allowed to occur
and could be described both in terms of wave and ray propagation.
With the present technology in manipulating the dielectric
coefficients in metamaterials \cite{smolyaninov, pendry_layered}, we hope that the effect
here derived can be experimentally tested and if verified could lead to applications.

\acknowledgments
This work was partially supported by the Brazilian research agencies
CNPq, FAPEMIG and CAPES. JPP thanks FAPEMIG for supporting his MSc studies
and Erasmus Mundus IRAP PhD Program for supporting his current PhD studies.


\begin{thebibliography}{88}
%
\bibitem{Landau}
L.D. Landau, E.M. Lifshitz, \emph{Electrodymanics of Continuous Media}, Vol. 8, Pergamon Press (1984).

\bibitem{born}
M. Born, and E. Wolf, \emph{Principles of Optics} (Cambridge
University Press, Cambridge, 1999).

\bibitem{paschotta2008}
R. Paschotta, { Encyclopedia of laser physics and technology}
(Wiley-VCH, Weinheim, 2008).

\bibitem{roth}
T. Roth and G.L.J.A. Rikken, { Phys. Rev. Lett.}, {\bf 85}, 4478 (2000); T. Roth and G.L.J.A. Rikken, {Phys. Rev. Lett.}, {\bf 88}, 063001 (2002).

\bibitem{delorenci2008}
V. A. De Lorenci and G. P. Goulart,
{ Phys. Rev. D}, {\bf 78} 045015 (2008).

\bibitem{bio}
L. Liu, J. R. Trimarchi, R. Oldenbourg, and D. L. Keefe,
{ Bio. Reprod.} {\bf 63}, 251 (2000);
L. Yiheng, {\it et al.}, {Biomed. Opt. Express} {\bf 2}, 2392 (2011).

\bibitem{astro1}
G. D. Fleishman, Q.J. Fu,  M. Wang, G.-L. Huang and V.F. Melnikov, {Phys. Rev. Lett.},  {\bf 88}, 251101
(2002); H.J.M. Cuesta, J.A. de Freitas Pacheco, J.M. Salim, {Int. J. Mod. Phys.
A}, {\bf 21}, 43 (2006).

\bibitem{astro2}
L. Pagano et al., {Phys. Rev. D}, {\bf 80}, 043522 (2009).

\bibitem{wood1969}
V. E. Wood and R. E. Mills,
{Optica Acta} {\bf 16}, 133 (1969).

\bibitem{tri1}
M. C. Netti, {\it et al.},
{Phys. Rev. Lett.} {\bf 86}, 1526 (2001).

\bibitem{Joannopoulos}
J.D. Joannopoulos, S.G. Johnson, J.N. Winn, R.D. Meade, \emph{Photonic Crystals: Molding the flow of light (Second Edition)}, (Princeton University Press, New Jersey, 2008).

\bibitem{smith2}
D. R. Smith, D. Schurig,
{Phys. Rev. Lett.}, {\bf 90}, 077405 (2003).

\bibitem{smith}
D. R. Smith, W. J. Padilla, D. C. Vier, S. C. Nemat-Nasser, and S. Schultz,
{Phys. Rev. Lett.} {\bf 84}, 4184 (2000).

\bibitem{veselago}
V. G. Veselago,
{Soviet Physics Uspekhi}, {\bf 10}, 509 (1968).

\bibitem{shelby}
R. A. Shelby, D. R. Smith and S. Schultz,
{Science}, {\bf 292}, 77 (2001).

\bibitem{trapping}
K. L. Tsakmakidis, A. D. Boardman and O. Hess,
{Nature}, {\bf 450}, 397 (2007);

\bibitem{smolyaninov3}
I. I. Smolyaninov, {\it et al.},
{Appl. Phys. Lett.}, {\bf 96}, 211121 (2010).

\bibitem{pendry2}
J. B. Pendry,
{Phys. Rev. Lett.}, {\bf 85}, 3966 (2000).

\bibitem{cloak1}
T. Ergin, N. Stenger, P. Brenner, J. B. Pendry, and M. Wegener,
{Science}, {\bf 328}, 337 (2010).

\bibitem{smolyaninov2}
I. I. Smolyaninov, V. N. Smolyaninova, A. V. Kildishev, and V. M. Shalaev,
{Phys. Rev. Lett.}, {102}, 213901 (2009).

\bibitem{asymmetric}
C. Menzel, {\it et al.},
{Phys. Rev. Lett.}, {\bf 104}, 253902 (2010).

\bibitem{smolyaninov}
I. I. Smolyaninov, E. E. Narimanov,
{Phys. Rev. Lett.}, {\bf 105}, 067402 (2010).

\bibitem{delorenci2010}
V. A. De Lorenci and D. D. Pereira,
{Phys. Rev.E}, {\bf 82} 036605 (2010).

\bibitem{Hadamard}
J. Hadamard, in {\it Le\c{c}ons sur la propagation
des ondes et les \'equations de l'hydrodynamique}
(Ed.\ Hermann, Paris, 1903);
G. Boillat, {J. Math.\ Phys.} {\bf 11}, 941 (1970);
A. Papapetrou, in {\it Lectures on General Relativity} (Dordrecht, Holland,1974).

\bibitem{delorenci2002}
V. A. De Lorenci and R. Klippert,
{Phys.\ Rev. D}, {\bf 65}, 064027 (2002); V. A. De Lorenci,
{Phys.\ Rev. E}, {\bf 65}, 026612 (2002).

\bibitem{delorenci2004}
V. A. De Lorenci, R. Klippert, and D. H. Teodoro,
{Phys. Rev. D }, {\bf 70}, 124035 (2004).

\bibitem{rodrigues}
R. R. Silva, J. Math. Phys. {\bf 39}, 6206 (1998).

\bibitem{smith3}
D. R. Smith, N. Kroll,
{Phys. Rev. Lett.}, {\bf 85}, 2933 (2000).

\bibitem{low_loss1}
A. Boltasseva and H. A. Atwater, {Science}, {\bf 331}, 290 (2011).

\bibitem{low_loss2}
P. Tassin, L. Zhang, Th. Koschny, E. N. Economou and C. M. Soukoulis, {Phys. Rev. Lett.}, {\bf 102}, 053901 (2009).

\bibitem{pendry_layered}
B. Wood, J. B. Pendry and D. P. Tsai, {Phys. Rev. B}, {\bf 74}, 115116 (2006).

\bibitem{baranova}
N. B. Baranova, Yu. V. Bogdanov and B. Ya. Zeldovich, {Usp. Fiz. Nauk.}, {\bf 123}, 349 (1977).
%
\end{thebibliography}
\end{document}